\documentclass{article}
\usepackage{times}
\newif\ifarxiv
\newif\ifcameraready

\camerareadyfalse
\arxivtrue

\ifcameraready
    \usepackage{iclr2027_conference}
    \iclrfinalcopy
\else
    \ifarxiv
        \usepackage{iclr2027_arxiv}
        \iclrfinalcopy
    \else
        \usepackage{iclr2027_conference}
    \fi
\fi

\usepackage{amsmath,amsfonts,bm}

\def\eqref#1{equation~\ref{#1}}

\def\1{\bm{1}}

\DeclareMathAlphabet{\mathsfit}{\encodingdefault}{\sfdefault}{m}{sl}
\SetMathAlphabet{\mathsfit}{bold}{\encodingdefault}{\sfdefault}{bx}{n}

\usepackage{algorithm}
\usepackage{algorithmic}
\usepackage{booktabs}
\usepackage{colortbl}
\usepackage{multirow}
\usepackage{graphicx}
\usepackage{placeins}
\usepackage{url}
\usepackage{hyperref}
\usepackage{subcaption}

\newcommand{\CCG}{\cellcolor{gray!15}}

\def\ours{\textsc{GTCW}}

\title{Grammar-Guided Code Watermarking with \\Green Temperature}

\author{Hyundong Jin, Hyeseon An, Soohan Lim, Yo-Sub Han\\
Department of Computer Science\\
Yonsei University, Seoul, Republic of Korea \\
\texttt{\{tuzi04,hsan,aness1219,emmous\}@yonsei.ac.kr}
}

\begin{document}

\maketitle

\begin{abstract}
Large language model watermarking embeds detectable statistical signals 
during decoding, but the resulting changes to 
token probabilities can degrade generation quality. 
This trade-off is particularly important for code, 
where small changes in token selection can break syntax 
or alter program behavior.
Existing code watermarking methods mitigate this risk 
through entropy-based insertion or syntax-aware token selection, 
but they do not directly construct the watermark 
over the set of continuations admitted by the current grammar state.
We propose Grammar-Guided Code Watermarking with Green Temperature~(\ours{}),
which integrates grammar-constrained decoding 
with probability-aware watermarking.
At each decoding step, \ours{} restricts the candidate set 
to grammar-admissible tokens and partitions this support 
into keyed green and red subsets.
At eligible high-entropy positions, green temperature reweights 
the green tokens according to the model's relative preferences,
strengthening the watermark signal while retaining the grammar constraint.
Across five models and five benchmarks spanning four programming languages,
\ours{} achieves a mean AUROC of $73.61\%$, 
compared with $67.83\%$ for the strongest baseline, 
while maintaining a mean Pass@1 of $59.18\%$ 
versus $59.58\%$ for unwatermarked generation.
Our implementation is available at \url{https://github.com/hyundong98/GTCW}.
\end{abstract}

\begin{figure}[!b]
\centering
\includegraphics[width=\linewidth]{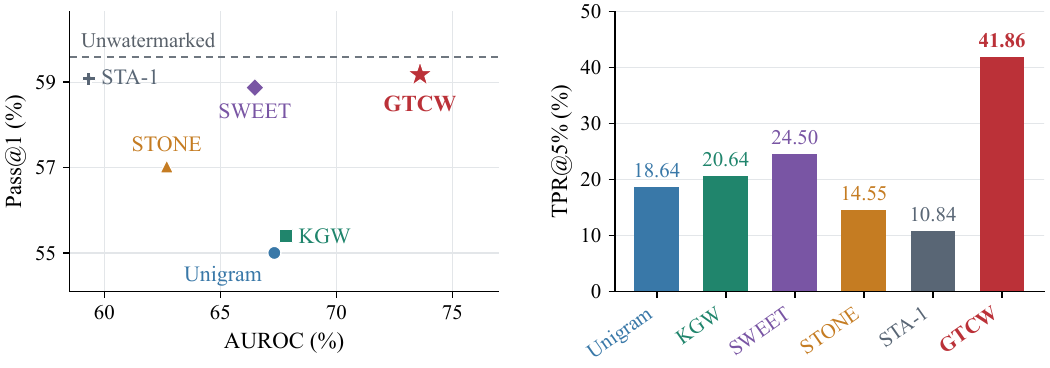}
\caption{AUROC-Pass@1 tradeoff (left) and TPR@5\%FPR (right), averaged equally
over five models, five benchmarks, and three seeds.
The star marks \ours{} and the dashed line denotes unwatermarked Pass@1
in the left panel.}
\label{GTCW:main:fig:overall_performance}
\end{figure}

\section{Introduction}
Large language models are increasingly used 
to generate executable code 
that can be directly incorporated into software 
systems~\citep{HouZLYWLLLGW2024}. 
Unlike transient model interactions, 
generated code may be copied, modified, and redistributed
independently of the model or service that produced it, 
making its provenance difficult to recover after generation. 
Reliable attribution of model-generated code is therefore 
becoming increasingly important 
as code generation is adopted more broadly in software development. 
Generation-time watermarking addresses this problem 
by embedding a detectable statistical signal 
into the generated code without retraining the underlying 
model~\citep{KirchenbauerGWKMG2023,ZhaoALW2024, DathathriSGHMWBKSMHVMBB2024}.

Many generation-time watermarks partition candidate tokens
into keyed green and red sets and modify the sampling distribution
in favor of the green set~\citep{KirchenbauerGWKMG2023,ZhaoALW2024,MaoWCFC2025, GuoTSLDL2024}.
For code, however, perturbing token probabilities can break syntax
or alter program behavior.
Existing code watermarking methods reduce this risk 
through entropy-based insertion or lexical and structural 
constraints~\citep{LeeHAHLYSK2024,KimPH2026,GuanWBWZZS2024,GuoC2025}.
These approaches improve the choice of watermark carriers,
but do not directly construct the watermark over the token set
admitted by the current grammar state.
A second source of distortion comes from 
how the selected tokens are reweighted.
Common green-list methods apply the same additive bias to all green tokens,
including those that the model originally assigns low probability~\citep{KirchenbauerGWKMG2023,LeeHAHLYSK2024,KimPH2026}.
In code generation, promoting such weak candidates can move probability mass
toward syntactically or functionally undesirable continuations 
rather than preserving the model's relative preference within the green 
set.

We propose Grammar-Guided Code Watermarking with Green Temperature~(\ours{})
to address both carrier selection and probability reweighting in code watermarking.
For carrier selection, grammar-constrained decoding restricts the candidate set
to admissible continuations, 
and \ours{} constructs the keyed green and red partition 
only within this support.
For probability reweighting, 
green temperature operates at eligible high-entropy positions 
and adjusts green tokens according to their model logits. 
This gives greater emphasis to candidates already preferred 
by the model rather than uniformly boosting all green tokens.
Detection replays the same grammar and keyed partitioning procedure on the
observed completion without requiring stored generation traces.

We evaluate \ours{} on five models and five benchmarks covering Python,
C++, Java, and Go. Across the 25 model-benchmark configurations,
\ours{} achieves a mean AUROC of $73.61\%$, compared with $67.83\%$ for the
strongest baseline, while its mean Pass@1 is $59.18\%$ compared with $59.58\%$
for unwatermarked generation. At TPR@5\%FPR, \ours{} reaches $41.86\%$,
compared with $24.50\%$ for the strongest baseline.
Figure~\ref{GTCW:main:fig:overall_performance} summarizes these aggregate results.

\section{Related Work}
\paragraph{LLM watermarking.}
Generation-time watermarking embeds a detectable statistical pattern
during decoding through logit modification 
or alternative sampling procedures~\citep{KirchenbauerGWKMG2023, ZhaoALW2024, 
KuditipudiTHL2024, DathathriSGHMWBKSMHVMBB2024, WangGWY2025}.
KGW introduced the soft green-list framework, 
which partitions candidate tokens into keyed green and red sets
while biasing generation toward the green subset~\citep{KirchenbauerGWKMG2023}.
Unigram watermarking uses a fixed keyed partition 
and provides formal robustness guarantees under editing and 
paraphrasing~\citep{ZhaoALW2024}. 
STA-1 introduces an unbiased sampling rule aimed at reducing distortion 
in low-entropy regimes~\citep{MaoWCFC2025}.
Unbiased and distortion-free watermarking methods more broadly seek to preserve
the original output distribution while retaining a detectable
signal~\citep{HuCWWZH2024,KuditipudiTHL2024, ChenWGH2025}.
These methods are directly applicable to pretrained models 
without code-specific training, 
which makes them natural references for our decoding-time setting, 
but they do not account for the grammatical structure
that restricts token choice in code.
Beyond additive green-list biasing, 
Permute-and-Flip develops a stable watermarkable decoder 
whose watermarking scheme preserves the underlying sampling
distribution~\citep{ZhaoLW2025}. 
This direction is related to our goal of reducing watermark-induced distortion,
but it redesigns the sampling rule itself and is not specialized to code 
or grammar-constrained carrier selection. 
We therefore treat it as a complementary general decoding approach
rather than a code-specific baseline.

\paragraph{Code watermarking.}
Code watermarking methods explicitly address 
the stronger structural constraints of program generation. 
SWEET restricts watermark scoring to high-entropy 
positions~\citep{LeeHAHLYSK2024}, 
while STONE avoids watermarking syntax-critical tokens 
and applies the watermark only to non-syntactic categories~\citep{KimPH2026}.
Both operate at decoding time without additional training,
making them direct baselines for our setting.
Other approaches incorporate learned structural information. 
CodeIP trains a type predictor and constrains next-token sampling 
according to predicted code-token types~\citep{GuanWBWZZS2024}.
ACW uses AST-aware training signals and semantic substitutability 
to adapt watermark insertion~\citep{GuoC2025}.
These approaches rely on learned or type-based structural guidance.
In contrast, \ours{} requires no auxiliary training and constructs a keyed
partition directly over the token set admitted by the current grammar state.
CodeGenGuard also uses training-based structural information,
but targets model ownership rather than output attribution~\citep{YangMXCZKJ2026}.
It encodes watermarks through semantics-preserving transformations
and verifies ownership 
using a trained watermark mechanism and trigger prompts.
It is therefore complementary rather than directly comparable
to our decoding-time output watermark.

\paragraph{Grammar-constrained generation.}
Programming languages are commonly described by context-free 
or closely related grammatical abstractions. 
This makes grammar constraints a natural way 
to restrict code generation to structurally valid continuations.
A line of work on grammar-constrained decoding enforces 
such constraints during LLM generation. 
SynCode uses CFG-based masking to restrict generation to tokens
consistent with a target grammar, 
while XGrammar improves structured-generation efficiency 
through optimized grammar execution and persistent parser-state
management~\citep{UgareSKMS2025,DongRCLXZC2025}.
In \ours{}, watermark construction is defined over the
state-dependent admissible support of the current grammar state,
and the keyed green partition is formed only over permitted tokens.

\section{Grammar-Guided Code Watermarking with Green Temperature}
\label{GTCW:main:sec:gtcw}

\ours{} addresses two sources of distortion in code watermarking:
the mismatch between vocabulary-level green partitions and the tokens
that are actually admissible under the current program state, 
and the probability distortion introduced 
by uniformly biasing selected green tokens.
We first formalize the former issue 
and then introduce grammar-guided partitioning, 
which constructs the watermark directly over the admissible token support. 
We next present green temperature for reweighting green candidates 
according to the model's relative preferences, 
followed by the corresponding replay-based detection procedure.

\subsection{Conditional Green-Rate Distortion}
\label{GTCW:main:ssec:green_distortion}
Let $x$ denote the prompt and $y=(y_1,\ldots,y_T)$ the generated continuation.
At decoding step $t$, $y_{<t}$ is the generated prefix 
and $\ell_t(v)$ is the model logit for token $v$.
The parser state $s_t$ determines the admissible token set
$\mathcal A_t$ under the current grammar state.
Conventional green-list watermarks construct a pseudo-random partition 
over the model vocabulary $\mathcal V$, 
with a nominal green fraction $\gamma$.
Let $\mathcal G_t\subseteq\mathcal V$ denote the green set 
at decoding step $t$.
For code generation, however, only a subset
$\mathcal A_t\subseteq\mathcal V$ is admissible 
at a given program state.
The effective green fraction among admissible tokens is therefore
\begin{equation}
\widehat{\gamma}_t
=
\frac{|\mathcal A_t\cap\mathcal G_t|}{|\mathcal A_t|}.
\label{eq:conditional-green-rate}
\end{equation}
If the green partition were sampled independently of $\mathcal A_t$,
we would expect $\mathbb E[\widehat{\gamma}_t]=\gamma$.
For a realized partition, however, $\widehat{\gamma}_t$ can deviate 
from the nominal green fraction.
This deviation is particularly relevant for context-dependent watermarks.
Their partition seeds are commonly derived from a recent token window, 
so repeated local token contexts can reproduce the same green-red partition.
Consequently, deviations in $\widehat{\gamma}_t$ may recur 
rather than average out across decoding steps.

Moreover, the same local token context can occur 
under different parser states with different admissible supports, 
while still inducing the same vocabulary-level partition.
For code watermarking, 
the relevant distribution is the one conditioned on tokens 
that can actually continue the current program.
This motivates constructing the green-red partition directly 
over the state-dependent admissible support, 
rather than partitioning the full vocabulary and conditioning on it afterward.

\subsection{Grammar-Guided Partitioning}
\label{GTCW:main:ssec:grammar_partition}

To maintain the target green fraction within the admissible support,
\ours{} constructs the green-red partition directly over $\mathcal A_t$.
Let $n_t=|\mathcal A_t|$, and let $c_t$ contain 
the most recent $w$ token identifiers.
The support signature $d_t$ is derived from the current admissible-token set.
A keyed hash $\mathcal H_k$ combines this signature 
with the local token context.
For $n_t>1$, we set
\begin{equation}
h_t=\mathcal H_k(d_t, c_t),
\qquad
m_t=\min\{n_t-1,\max\{1,\lfloor\gamma n_t+\tfrac12\rfloor\}\}.
\label{GTCW:main:eq:state_seed}
\end{equation}
The seed $h_t$ induces a pseudo-random ranking of $\mathcal A_t$, 
and the first $m_t$ tokens form the green set $\mathcal G_t$.
The remaining tokens are red.
This choice keeps both sets nonempty 
while maintaining the green fraction $\gamma_t=m_t/n_t$ 
close to the target $\gamma$.
When $n_t\leq1$, we set $\mathcal G_t=\emptyset$ and insert no watermark.

Including $d_t$ in the seed makes green membership depend on 
both the recent token context and the current admissible support.
The same local context can therefore induce different green-red partitions
under different grammar states.
As a result, a partition associated with one admissible support is not
directly reusable under another.
States with the same support signature and token context deterministically
share the same partition.

\begin{figure}[t]
\centering
\includegraphics[width=\linewidth]{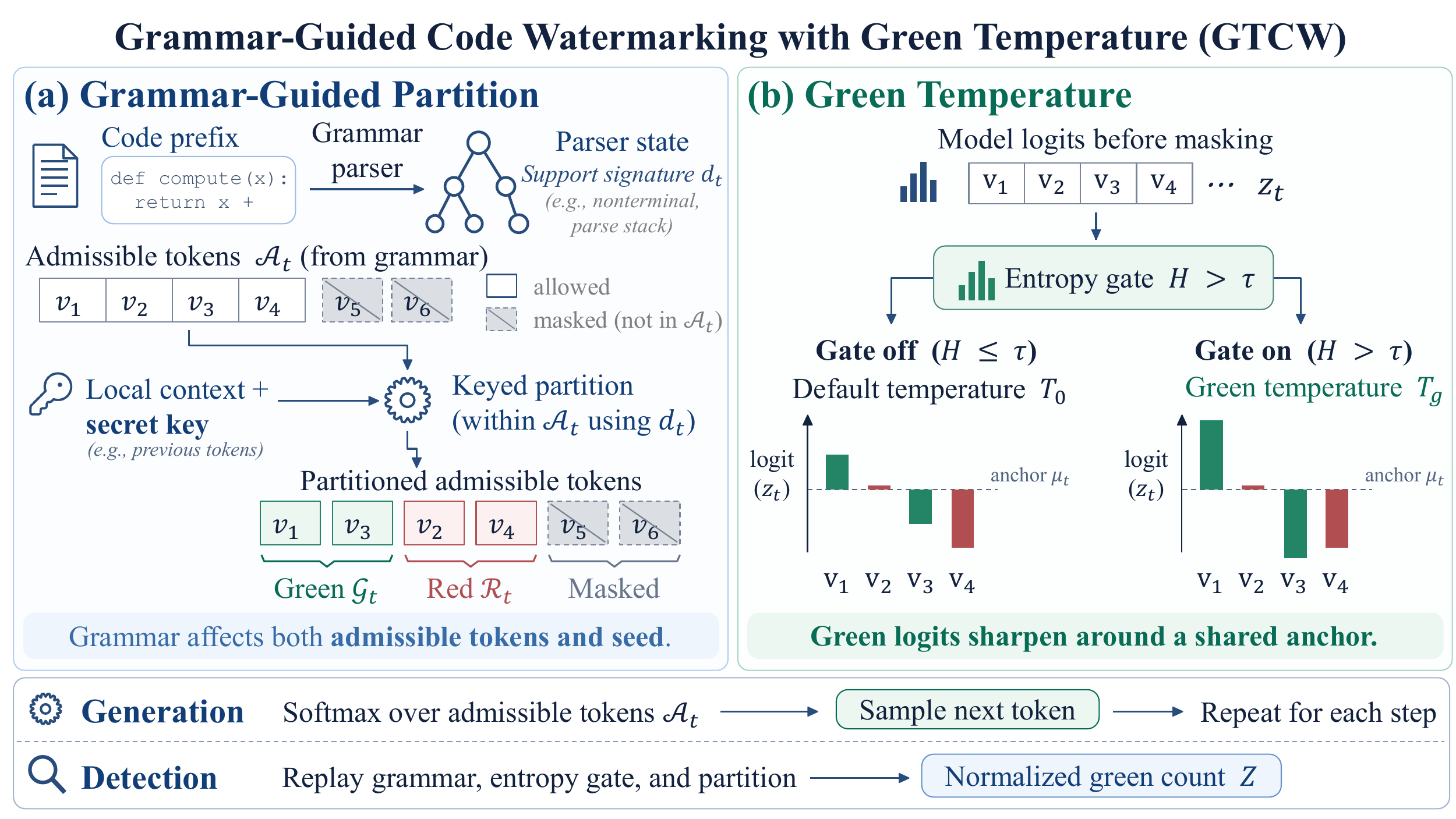}
\caption{
Overview of \ours{}.
(a) Grammar restricts the token support and contributes to keyed partitioning
over admissible tokens.
(b) At entropy-eligible positions, green temperature sharpens green logits
around a shared anchor.
}
\label{GTCW:main:fig:method-overview}
\end{figure}

\subsection{Green Temperature}
\label{GTCW:main:ssec:gtcw-biasing}

After constructing the green-red partition, 
the remaining question is how to reweight the selected green tokens.
A standard additive bias shifts all green logits by the same amount, 
regardless of their original model probabilities.
Instead, green temperature adjusts green tokens 
according to their relative model preferences, 
while leaving red logits unchanged within the admissible support.

We apply the watermark only at sufficiently uncertain decoding steps.
Let $p_t=\operatorname{softmax}(\ell_t)$ denote the model distribution
before grammar masking or watermarking.
With entropy threshold $\tau$, we define
\begin{equation}
H_t=-\sum_{v\in\mathcal V}p_t(v)\log p_t(v),
\qquad
a_t=e_t\,\mathbf 1[H_t>\tau]\,\mathbf 1[n_t>1].
\label{eq:entropy-gate}
\end{equation}
Here $e_t$ is a binary carrier gate that suppresses watermark insertion
during long comment or whitespace runs in C++ and Java.
The grammar constraint remains active even when $a_t=0$.

Let $T_0$ be the default sampling temperature and 
$T_g<T_0$ the green temperature, and define $\beta=1/T_g-1/T_0$.
We center the adjustment around the log-mean-exp of the admissible logits,
\begin{equation}
\mu_t=
\log\left(
\frac{1}{|\mathcal A_t|}
\sum_{v\in\mathcal A_t}\exp(\ell_t(v))
\right).
\label{eq:logit-anchor}
\end{equation}
The adjusted logits are
\begin{equation}
\widetilde\ell_t(v)=
\begin{cases}
\ell_t(v)/T_0+
a_t\beta\bigl(\ell_t(v)-\mu_t\bigr)
\mathbf 1[v\in\mathcal G_t],
& v\in\mathcal A_t,\\
-\infty,
& v\notin\mathcal A_t.
\end{cases}
\label{eq:green-logits}
\end{equation}
Conditioned on generating a source token, 
we sample from $q_t(v)\propto\exp(\widetilde\ell_t(v))$.
Termination and control tokens are handled separately 
from this source-token distribution.
At an eligible position, 
the adjusted logit of a green token can equivalently be written as
\begin{equation}
\frac{\mu_t}{T_0}
+
\frac{\ell_t(v)-\mu_t}{T_g}.
\end{equation}

Green temperature preserves the ordering among green tokens 
while sharpening their relative preferences.
For two green tokens $u,v\in\mathcal A_t$ at an eligible position,
\begin{equation}
\frac{q_t(u)}{q_t(v)}
=
\exp\left(
\frac{\ell_t(u)-\ell_t(v)}{T_g}
\right).
\label{eq:green-ratio}
\end{equation}
Thus, green tokens already preferred by the model receive greater 
relative emphasis, rather than all green tokens receiving the same boost.
In contrast, a uniform additive bias changes the probability mass 
assigned to the green set 
while leaving the relative distribution within that set unchanged.
The anchor makes the transformation invariant to a common shift of the logits.

\subsection{Detection}
\label{GTCW:main:ssec:gtcw}

Detection takes the completed code, 
separating the known code prefix from the generated continuation, 
and replays the generation states 
using the original model, grammar, and secret key.
Before consuming token $y_t$, the detector reconstructs 
the admissible set $\mathcal A_t$, watermark eligibility $a_t$, 
and green set $\mathcal G_t$.
We score the eligible positions
\begin{equation}
\mathcal T = \{t \mid a_t=1,\ y_t\in\mathcal A_t\}.
\end{equation}
For each $t\in\mathcal T$, let
$I_t=\mathbf 1[y_t\in\mathcal G_t]$ and
$\gamma_t=m_t/n_t$.
Under an idealized random-partition null,
$I_t$ has mean $\gamma_t$ and variance
$\gamma_t(1-\gamma_t)$.
We therefore use the normalized green count
\begin{equation}
Z=
\frac{
\sum_{t\in\mathcal T}(I_t-\gamma_t)
}{
\sqrt{
\sum_{t\in\mathcal T}\gamma_t(1-\gamma_t)
}
}.
\label{eq:detection-score}
\end{equation}

A larger $Z$ indicates stronger evidence that the continuation follows the
keyed green partitions.
If an observed token is inadmissible or parsing fails, replay stops and retains
the evidence accumulated over the admissible prefix.
If no eligible position is available, no finite detection score is produced,
and the output is retained in evaluation as having no watermark evidence.
Detection requires model inference to reconstruct the entropy gate, but does
not require stored generation traces.

\section{Experiments}

\subsection{Experimental Settings}
\label{GTCW:main:ssec:experimental_settings}

We evaluate \ours{} across five instruction-tuned code generation models
and five benchmarks covering Python, C++, Java, and Go, including HumanEval+,
MBPP+, and multilingual subsets of HumanEvalPack, 
with functional correctness measured by execution-based 
tests~\citep{LiuXWZ2023,MuennighoffLZZHZSTWL2024}.
We compare against Unigram, KGW, SWEET, STONE, and STA-1, 
together with an unwatermarked Base reference, 
using matched prompts and sampling configurations
for each model-benchmark setting.
We report AUROC and TPR@5\%FPR for watermark detection and Pass@1 
for functional correctness, 
retaining all generated samples and comparing watermarked outputs 
against matched unwatermarked completions.
Results are averaged over three random seeds 
with equal weight across model-benchmark configurations.
The main table reports Qwen2.5-Coder~(3B/7B) and Llama3.1~(8B), 
while the complete five-model results are 
provided in Appendix~\ref{GTCW:app:ssec:full_main_results}.
Sensitivity, ablation, and code-transformation evaluations use
Qwen2.5-Coder~(3B).
The transformation evaluation includes identifier renaming, formatting,
comment removal, dead-code insertion, control-flow transformations,
and LLM rewriting.
Detailed experimental settings and metric definitions are 
provided in Appendix~\ref{GTCW:app:sec:experimental_details}.

\subsection{Main Results}
\label{GTCW:main:ssec:main_results}

\begin{table}[t]
\centering
\caption{
Detection and functional correctness.
AUC, T@5, and P@1 denote AUROC, TPR@5\%FPR, and Pass@1, respectively.
Values are mean percentages over three seeds.
Gray shading indicates \ours{}.
Bold and underline mark the best and second-best watermarked methods,
respectively, based on unrounded values.
}
\label{GTCW:main:tab:main_results}
\small
\setlength{\tabcolsep}{1.5pt}
\begin{tabular}{@{}cc*{5}{rrr}@{}}
\toprule[1.2pt]
& & \multicolumn{3}{c}{HumanEval+} & \multicolumn{3}{c}{MBPP+} & \multicolumn{9}{c}{HumanEvalPack} \\
\cmidrule(lr){3-5}\cmidrule(lr){6-8}\cmidrule(lr){9-17}
Model & Method & \multicolumn{3}{c}{Python} & \multicolumn{3}{c}{Python} & \multicolumn{3}{c}{C++} & \multicolumn{3}{c}{Java} & \multicolumn{3}{c}{Go} \\
\cmidrule(lr){3-5}\cmidrule(lr){6-8}\cmidrule(lr){9-11}\cmidrule(lr){12-14}\cmidrule(lr){15-17}
& & AUC & T@5 & P@1 & AUC & T@5 & P@1 & AUC & T@5 & P@1 & AUC & T@5 & P@1 & AUC & T@5 & P@1 \\
\midrule[1.2pt]
\multirow{7}{*}{\shortstack{Qwen2.5-C\\(3B)}} & Base & --- & --- & 67.9 & --- & --- & 52.1 & --- & --- & 52.2 & --- & --- & 71.8 & --- & --- & 51.2 \\
 & Unigram & 68.1 & 18.9 & 65.0 & 82.4 & 38.5 & 43.5 & 75.5 & 28.5 & 46.8 & 67.2 & 16.9 & 63.4 & 71.2 & 15.5 & 45.7 \\
 & KGW & 68.1 & 22.0 & 63.4 & 81.2 & 39.7 & 47.6 & \underline{76.7} & \underline{33.9} & 45.3 & 65.5 & 14.4 & 66.9 & 68.6 & 22.0 & 47.6 \\
 & SWEET & \underline{68.9} & \underline{24.4} & 68.7 & \underline{84.4} & \underline{41.8} & 50.8 & 71.0 & 30.1 & \textbf{53.9} & \underline{68.5} & \underline{25.8} & \textbf{71.1} & \underline{72.8} & \underline{29.3} & 48.2 \\
 & STONE & 63.3 & 11.2 & 67.1 & 74.3 & 24.1 & 47.1 & 66.4 & 16.3 & 49.2 & 61.8 & 11.4 & \underline{70.3} & 63.7 & 16.1 & 46.5 \\
 & STA-1 & 57.9 & 7.9 & \textbf{70.5} & 70.5 & 20.0 & \underline{51.2} & 62.8 & 11.2 & \underline{51.0} & 59.8 & 11.4 & 69.3 & 60.5 & 8.9 & \textbf{51.8} \\
 & \CCG \ours{} & \CCG \textbf{77.6} & \CCG \textbf{49.2} & \CCG \underline{69.5} & \CCG \textbf{96.5} & \CCG \textbf{83.6} & \CCG \textbf{51.8} & \CCG \textbf{81.1} & \CCG \textbf{51.4} & \CCG 50.0 & \CCG \textbf{75.9} & \CCG \textbf{37.2} & \CCG \underline{70.3} & \CCG \textbf{86.4} & \CCG \textbf{56.5} & \CCG \underline{49.2} \\
\midrule
\multirow{7}{*}{\shortstack{Qwen2.5-C\\(7B)}} & Base & --- & --- & 77.0 & --- & --- & 61.3 & --- & --- & 63.6 & --- & --- & 79.3 & --- & --- & 63.2 \\
 & Unigram & 60.5 & \underline{12.2} & 74.4 & \underline{80.6} & 32.6 & 53.1 & \underline{67.3} & \underline{17.3} & 60.4 & \underline{59.2} & 9.4 & \underline{78.9} & \textbf{62.7} & 10.2 & 61.8 \\
 & KGW & \underline{60.6} & 7.9 & 75.2 & 79.4 & \underline{40.0} & 55.0 & \textbf{67.9} & 15.5 & 56.9 & \textbf{60.9} & 8.7 & 74.6 & 60.6 & \underline{11.8} & 59.4 \\
 & SWEET & 57.0 & 3.9 & 76.8 & 75.4 & 27.3 & \underline{60.0} & 59.1 & 11.6 & \underline{62.6} & 53.8 & \underline{11.2} & 77.2 & 57.7 & 8.3 & \textbf{63.0} \\
 & STONE & 59.3 & 10.8 & 76.6 & 69.8 & 20.9 & 57.6 & 58.7 & 7.5 & 62.0 & 55.7 & 6.7 & 77.6 & 58.2 & 10.8 & \textbf{63.0} \\
 & STA-1 & 57.3 & 10.0 & \textbf{77.2} & 65.6 & 15.8 & \textbf{60.9} & 58.6 & 8.1 & \textbf{63.2} & 55.1 & 7.9 & 76.0 & 55.6 & 7.3 & 60.2 \\
 & \CCG \ours{} & \CCG \textbf{60.9} & \CCG \textbf{21.5} & \CCG \underline{77.0} & \CCG \textbf{88.3} & \CCG \textbf{56.0} & \CCG 59.5 & \CCG 64.8 & \CCG \textbf{27.4} & \CCG 58.1 & \CCG 56.7 & \CCG \textbf{16.1} & \CCG \textbf{80.1} & \CCG \underline{62.6} & \CCG \textbf{20.9} & \CCG \underline{62.2} \\
\midrule
\multirow{7}{*}{\shortstack{Llama3.1\\(8B)}} & Base & --- & --- & 50.6 & --- & --- & 57.8 & --- & --- & 40.0 & --- & --- & 53.1 & --- & --- & 32.9 \\
 & Unigram & \underline{70.2} & 21.5 & 46.8 & \underline{63.4} & 12.3 & 56.4 & 76.7 & 24.6 & 34.2 & 70.2 & 22.4 & 45.5 & 71.0 & 26.0 & 23.8 \\
 & KGW & 66.6 & 23.0 & 48.4 & 63.0 & \underline{15.5} & 55.8 & 76.7 & 31.3 & 33.5 & 70.3 & 25.2 & 47.6 & 73.9 & 25.6 & 30.9 \\
 & SWEET & 69.9 & \underline{27.2} & 49.4 & 62.1 & 15.1 & 57.0 & \underline{81.2} & \underline{41.5} & \underline{36.8} & \underline{73.3} & \underline{33.3} & 51.6 & \underline{82.6} & \underline{47.0} & 30.5 \\
 & STONE & 65.5 & 16.9 & 48.0 & 60.6 & 12.9 & 56.4 & 69.6 & 21.5 & \underline{36.8} & 63.4 & 13.0 & 51.4 & 69.0 & 22.0 & 29.5 \\
 & STA-1 & 55.7 & 10.4 & \underline{51.4} & 56.0 & 9.0 & \underline{57.6} & 64.8 & 14.6 & \textbf{39.2} & 59.1 & 17.1 & \underline{52.0} & 62.1 & 13.4 & \textbf{36.0} \\
 & \CCG \ours{} & \CCG \textbf{78.6} & \CCG \textbf{45.9} & \CCG \textbf{53.1} & \CCG \textbf{66.6} & \CCG \textbf{27.7} & \CCG \textbf{58.6} & \CCG \textbf{90.5} & \CCG \textbf{76.2} & \CCG 33.7 & \CCG \textbf{79.6} & \CCG \textbf{47.0} & \CCG \textbf{52.4} & \CCG \textbf{92.8} & \CCG \textbf{73.0} & \CCG \underline{31.5} \\
\bottomrule[1.2pt]
\end{tabular}%
\end{table}

Table~\ref{GTCW:main:tab:main_results} compares the three models 
across five benchmarks.
Across these 15 model-benchmark settings, 
\ours{} achieves the highest TPR@5\%FPR in every setting 
and the highest AUROC in 12 of 15 settings.
The advantage is particularly pronounced 
on Qwen2.5-Coder~(3B) and Llama3.1~(8B), 
while Qwen2.5-Coder~(7B) shows a smaller AUROC margin 
but retains the same consistent advantage at the low-FPR operating point.
At the same time, \ours{} remains competitive in Pass@1 
across the benchmarks, 
showing that the stronger detection is achieved 
without substantially degrading functional correctness.
Appendix~\ref{GTCW:app:ssec:full_main_results} extends the comparison to
Qwen3.5~(9B) and Gemma4~(12B).
Across all five models and five benchmarks, 
\ours{} achieves mean AUROC of $73.61\%$ and TPR@5\%FPR of $41.86\%$, 
improving over the strongest baselines 
by $5.78$ and $17.36$ percentage points, respectively.
Its mean Pass@1 is $59.18\%$, compared with $59.58\%$ for Base.

\paragraph{Variation across models.}
The magnitude of the detection gain varies noticeably across models.
\ours{} is particularly effective on Qwen2.5-Coder~(3B) and Llama3.1~(8B), 
while the AUROC margin becomes smaller on Qwen2.5-Coder~(7B).
This variation appears to be related less to parameter count itself than
to how each model distributes probability over watermarkable positions.
Green temperature is active only 
when the model retains sufficient uncertainty 
among grammar-admissible candidates, 
so a model that produces sharper distributions can provide fewer opportunities 
to accumulate watermark evidence.
This interpretation is consistent with Qwen2.5-Coder, 
where finite-score coverage drops from $88.48\%$ at 3B to $66.51\%$ at 7B 
despite the two checkpoints belonging to the same code-specialized family.
Even in the 7B setting, however, \ours{} retains the highest
TPR@5\%FPR on all five benchmarks, suggesting that the selected carrier
positions remain effective when available.

\paragraph{Variation across benchmarks.}
The benchmark differences reflect a different interaction 
between grammar constraints and the semantic requirements of the generated code.
MBPP+ combines strong detection with 
almost unchanged Pass@1 relative to Base, 
suggesting that many of its watermarkable decisions can be
reweighted without changing the final program behavior.
Go similarly provides strong detection, 
although with a somewhat larger correctness decrease.

C++ shows a different pattern.
Although \ours{} maintains high detection performance, 
it exhibits the largest mean Pass@1 decrease among the evaluated benchmarks.
Our grammar filters token choices according to syntactic admissibility,
but it does not enforce type consistency, API usage, or program semantics.
This distinction is particularly relevant for C++, 
where a continuation can remain grammatically admissible 
while a small token-level change still affects compilation or execution behavior.
The result therefore highlights the role of grammar guidance 
as a protective constraint rather than a complete correctness guarantee.
Despite these language-dependent effects, \ours{} preserves 
aggregate functional correctness close to Base 
while maintaining substantially stronger detection.
The following ablation further separates the roles of grammar guidance
and green temperature in this trade-off.

\subsection{Ablation Study}

Table~\ref{GTCW:main:tab:ablation} separates the two central design
choices of \ours{}: restricting watermarking through grammar guidance
and replacing uniform additive bias with green temperature.
The results show complementary roles, 
with grammar guidance primarily protecting functional correctness 
and green temperature providing the larger gain in detectability.

\begin{table}[ht]
\centering
\caption{
Ablation of grammar guidance and green temperature on
Qwen2.5-Coder~(3B).
All settings use the same entropy gate.
AUC, T@5, and P@1 denote AUROC, TPR@5\%FPR, and Pass@1.
Values are mean percentages over three seeds.
}
\label{GTCW:main:tab:ablation}
\small
\setlength{\tabcolsep}{0.85pt}
\begin{tabular}{@{}c*{18}{r}@{}}
\toprule[1.2pt]
 & \multicolumn{3}{c}{HumanEval+}
 & \multicolumn{3}{c}{MBPP+}
 & \multicolumn{9}{c}{HumanEvalPack}
 & \multicolumn{3}{c}{\multirow{2}{*}[-2pt]{AVG}} \\
\cmidrule(lr){2-4}\cmidrule(lr){5-7}\cmidrule(lr){8-16}
Setting
 & \multicolumn{3}{c}{Python}
 & \multicolumn{3}{c}{Python}
 & \multicolumn{3}{c}{C++}
 & \multicolumn{3}{c}{Java}
 & \multicolumn{3}{c}{Go}
 & & & \\
\cmidrule(lr){2-4}\cmidrule(lr){5-7}\cmidrule(lr){8-10}
\cmidrule(lr){11-13}\cmidrule(lr){14-16}\cmidrule(lr){17-19}
 & AUC & T@5 & P@1
 & AUC & T@5 & P@1
 & AUC & T@5 & P@1
 & AUC & T@5 & P@1
 & AUC & T@5 & P@1
 & AUC & T@5 & P@1 \\
\midrule[1.2pt]
w/o Gram.
 & \underline{76.4} & \underline{36.4} & 63.2
 & \underline{96.0} & \underline{76.5} & \textbf{52.2}
 & \textbf{83.1} & \textbf{57.3} & 49.0
 & \textbf{76.4} & \textbf{40.7} & \underline{69.7}
 & \underline{85.0} & \textbf{60.0} & 45.5
 & \underline{83.4} & \underline{54.2} & 55.9 \\
w/o $T_g$
 & 68.7 & 27.0 & \textbf{69.7}
 & 86.8 & 50.4 & 50.0
 & 72.3 & 35.4 & \underline{49.4}
 & 67.8 & 21.3 & 68.9
 & 76.7 & 29.9 & \textbf{50.0}
 & 74.5 & 32.8 & \underline{57.6} \\
\CCG \ours{}
 & \CCG \textbf{77.6} & \CCG \textbf{49.2} & \CCG \underline{69.5}
 & \CCG \textbf{96.5} & \CCG \textbf{83.6} & \CCG \underline{51.8}
 & \CCG \underline{81.1} & \CCG \underline{51.4} & \CCG \textbf{50.0}
 & \CCG \underline{75.9} & \CCG \underline{37.2} & \CCG \textbf{70.3}
 & \CCG \textbf{86.4} & \CCG \underline{56.5} & \CCG \underline{49.2}
 & \CCG \textbf{83.5} & \CCG \textbf{55.6} & \CCG \textbf{58.2} \\
\bottomrule[1.2pt]
\end{tabular}
\end{table}

\paragraph{Grammar guidance.}
Grammar guidance primarily improves functional correctness.
Removing it lowers mean Pass@1 from $58.16\%$ to $55.93\%$, while
mean AUROC changes only from $83.51\%$ to $83.38\%$.
The correctness gain is especially large on HumanEval+ and Go, where
grammar guidance improves Pass@1 by $6.30$ and $3.66$ points,
respectively.
This behavior matches the intended role of grammar guidance.
By restricting watermarking to tokens admitted by the current grammar
state, \ours{} avoids shifting probability toward syntactically
inadmissible continuations while retaining opportunities to embed the
watermark.
The detection effect is more benchmark-dependent, but the aggregate
AUROC remains essentially unchanged while Pass@1 improves by $2.23$ points.
Thus, grammar guidance acts primarily as a quality-preserving mechanism
rather than the main source of watermark strength.

\paragraph{Green temperature.}
Green temperature provides the larger gain in watermark detectability.
With grammar guidance and the entropy gate fixed, replacing green
temperature with additive bias reduces mean AUROC from $83.51\%$ to
$74.46\%$ and mean TPR@5\%FPR from $55.6\%$ to $32.8\%$.
\ours{} improves both detection metrics on all five benchmarks, showing
that the advantage is consistent across Python, C++, Java, and Go.
Importantly, this stronger signal does not come at the expense of average
functional correctness.
Mean Pass@1 increases slightly from $57.60\%$ with additive bias to
$58.16\%$ with green temperature.
This supports the motivation for green temperature. 
Rather than uniformly boosting every green token, 
it uses the model's existing preferences to favor stronger green candidates.
Together with the grammar ablation, 
these results show a clear division of labor between the two components.

\begin{table}[th]
\centering
\setlength{\belowcaptionskip}{6pt}
\caption{
Sensitivity to green temperature $T_g$ on Qwen2.5-Coder~(3B) with $T_0=1$.
AUC, T@5, and P@1 denote AUROC, TPR@5\%FPR, and Pass@1.
Values are mean percentages over three seeds.
Gray shading indicates the default $T_g=0.5$.
}
\label{GTCW:main:tab:gt_sensitivity}
\label{tab:gt-sensitivity}
\small
\setlength{\tabcolsep}{3.5pt}
\begin{tabular}{@{}c*{15}{r}@{}}
\toprule[1.2pt]
 & \multicolumn{3}{c}{HumanEval+}
 & \multicolumn{3}{c}{MBPP+}
 & \multicolumn{9}{c}{HumanEvalPack} \\
\cmidrule(lr){2-4}\cmidrule(lr){5-7}\cmidrule(lr){8-16}
$T_g$ & \multicolumn{3}{c}{Python}
 & \multicolumn{3}{c}{Python}
 & \multicolumn{3}{c}{C++}
 & \multicolumn{3}{c}{Java}
 & \multicolumn{3}{c}{Go} \\
\cmidrule(lr){2-4}\cmidrule(lr){5-7}\cmidrule(lr){8-10}
\cmidrule(lr){11-13}\cmidrule(lr){14-16}
 & AUC & T@5 & P@1
 & AUC & T@5 & P@1
 & AUC & T@5 & P@1
 & AUC & T@5 & P@1
 & AUC & T@5 & P@1 \\
\midrule[1.2pt]
$0.3$
 & \textbf{78.0} & \textbf{51.8} & 69.3
 & \textbf{96.7} & \textbf{84.9} & 52.4
 & \textbf{82.9} & \textbf{54.3} & 47.8
 & \textbf{76.6} & \textbf{38.4} & 70.1
 & \textbf{87.1} & \textbf{59.1} & 48.6 \\
$0.4$
 & 77.9 & 50.8 & 69.3
 & \textbf{96.7} & 84.3 & 52.1
 & 82.2 & 52.6 & 49.0
 & 76.3 & 38.2 & \textbf{70.5}
 & 86.9 & 58.9 & 48.0 \\
\CCG $0.5$
 & \CCG 77.6 & \CCG 49.2 & \CCG \textbf{69.5}
 & \CCG 96.5 & \CCG 83.6 & \CCG 51.8
 & \CCG 81.1 & \CCG 51.4 & \CCG 50.0
 & \CCG 75.9 & \CCG 37.2 & \CCG 70.3
 & \CCG 86.4 & \CCG 56.5 & \CCG \textbf{49.2} \\
$0.6$
 & 76.4 & 46.5 & 68.7
 & 95.6 & 81.0 & 51.5
 & 79.1 & 47.6 & 49.8
 & 74.3 & 34.3 & 69.9
 & 84.7 & 51.4 & 49.0 \\
$0.7$
 & 73.8 & 38.6 & 68.7
 & 93.3 & 72.2 & \textbf{52.6}
 & 76.7 & 44.7 & \textbf{50.2}
 & 72.1 & 28.3 & 70.3
 & 82.0 & 43.5 & 48.8 \\
\bottomrule[1.2pt]
\end{tabular}
\end{table}

\subsection{Green Temperature Sensitivity}

Table~\ref{GTCW:main:tab:gt_sensitivity} varies $T_g$ from $0.3$ to $0.7$
while keeping all other settings fixed.
Lower green temperatures consistently strengthen detection.
As $T_g$ approaches the base temperature $T_0=1$, the green-temperature
adjustment becomes weaker, 
and both AUC and T@5 decrease across all five benchmarks.
Mean AUROC falls from $84.25\%$ at $T_g=0.3$ to $79.58\%$ at $0.7$,
while mean TPR@5\%FPR decreases from $57.7\%$ to $45.5\%$.
This trend follows directly from the green-temperature transformation.
As $T_g$ decreases, probability becomes increasingly concentrated on
green candidates already assigned high logits by the model, yielding a
stronger watermark signal.

At the same time, overall performance is not overly sensitive to moderate
changes in $T_g$.
In particular, detection remains relatively stable around the default.
Mean AUROC differs by only $0.75$ points between $T_g=0.3$ and $0.5$,
with the larger decline appearing as $T_g$ moves closer to $T_0$.
Functional correctness is even less sensitive 
and does not follow a monotonic trend.
Across the displayed range, mean Pass@1 varies only 
from $57.63\%$ to $58.16\%$.
This stability suggests that stronger multiplicative reweighting need not
directly translate into lower code quality.
Since green temperature sharpens preferences among already favored
candidates rather than uniformly promoting weak green tokens, a stronger
watermark can remain compatible with the concentrated token choices that
are often effective for code generation.
The default $T_g=0.5$ attains the highest mean Pass@1 while remaining close
to the strongest detection settings.
We therefore use $T_g=0.5$ as a balanced operating point between watermark
strength and functional correctness.

\subsection{Detection after Code Transformations}
\label{GTCW:main:ssec:transformation_detection}

We evaluate detection after six transformation families:
identifier renaming, formatting, comment removal, dead-code insertion,
control-flow transformations, and LLM rewriting.
Table~\ref{GTCW:main:tab:transformation_detection} reports detection after each
transformation, with the clean outputs shown as a reference.
Across the 13 transformation settings, \ours{} achieves the highest
average TPR@5\%FPR at $27.2\%$ and ranks first or second in T@5 in
every setting.
Its average AUROC is $70.1\%$, second to Unigram at $73.2\%$.
This mirrors the main evaluation: the most consistent advantage of
\ours{} appears at the low-FPR operating point, even when overall AUROC
varies across transformations.
Details of the transformation protocol are provided in
Appendix~\ref{GTCW:app:ssec:transformation_protocol}.

\paragraph{Local code transformations.}
\ours{} remains particularly strong under comment removal and control-flow
changes.
It achieves the highest AUC and T@5 after comment removal and under all
three control-flow settings.
The comment-removal result is notable because it shows that the detectable
signal is not concentrated in comments alone.
Likewise, the control-flow results indicate that local structural changes
can preserve substantial watermark evidence even when the generated syntax
is reorganized.
Identifier renaming and formatting have a larger effect.
Under renaming, \ours{} has the highest T@5 at $25\%$ and $50\%$ strength
and remains second at $75\%$ and $100\%$.
After formatting, its AUROC decreases to $67.8\%$, but it still retains
the highest T@5 at $27.5\%$.
These results suggest that the watermark is not tied to a single
lexical feature, although changes to the token sequence 
and local grammar state can weaken replayed evidence.

\begin{table}[th]
\centering
\caption{
Detection after code transformations on Qwen2.5-Coder~(3B).
AUC and T@5 denote AUROC and TPR@5\%FPR.
Values average five benchmarks and three seeds.
AVG averages the 13 transformation settings, excluding Clean.
Gray shading indicates \ours{}, with bold and underline marking 
the best and second-best methods.
}
\label{GTCW:main:tab:transformation_detection}
\small
\setlength{\tabcolsep}{2pt}
\begin{tabular}{@{}ll*{12}{r}@{}}
\toprule[1.2pt]
\cmidrule(lr){3-14}
\multirow{2}{*}[-2pt]{Transformation} & \multirow{2}{*}[-2pt]{Setting} & \multicolumn{2}{c}{Unigram}
 & \multicolumn{2}{c}{KGW}
 & \multicolumn{2}{c}{SWEET}
 & \multicolumn{2}{c}{STONE}
 & \multicolumn{2}{c}{STA-1}
 & \multicolumn{2}{c}{\ours{}} \\
\cmidrule(lr){3-4}\cmidrule(lr){5-6}\cmidrule(lr){7-8}
\cmidrule(lr){9-10}\cmidrule(lr){11-12}\cmidrule(lr){13-14}
 & & AUC & T@5 & AUC & T@5 & AUC & T@5
 & AUC & T@5 & AUC & T@5 & AUC & T@5 \\
\midrule[1.2pt]
Clean & ---
 & 72.9 & 23.6
 & 72.0 & 26.4
 & \underline{73.1} & \underline{30.3}
 & 65.9 & 15.8
 & 62.3 & 11.9
 & \CCG \textbf{83.5} & \CCG \textbf{55.6} \\
\midrule
\multirow{4}{*}{Renaming} & 25\%
 & \textbf{73.5} & \underline{24.6}
 & 67.3 & 24.4
 & 72.0 & 23.3
 & 64.5 & 17.3
 & 60.6 & 12.0
 & \CCG \underline{73.2} & \CCG \textbf{29.2} \\
 & 50\%
 & \textbf{74.3} & \underline{26.1}
 & 66.9 & 23.7
 & 71.3 & 22.7
 & 63.7 & 17.8
 & 59.5 & 12.0
 & \CCG \underline{72.4} & \CCG \textbf{27.8} \\
 & 75\%
 & \textbf{75.1} & \textbf{26.3}
 & 65.0 & 22.6
 & \underline{70.8} & 21.6
 & 63.1 & 17.7
 & 59.1 & 11.8
 & \CCG 69.6 & \CCG \underline{25.0} \\
 & 100\%
 & \textbf{75.4} & \textbf{26.4}
 & 64.0 & 22.4
 & \underline{70.0} & 21.3
 & 62.9 & 17.5
 & 58.1 & 11.2
 & \CCG 69.4 & \CCG \underline{24.6} \\
\midrule
Formatting & Normalized
 & \textbf{73.3} & \underline{26.3}
 & \underline{69.1} & 24.4
 & 67.7 & 22.3
 & 68.3 & 17.9
 & 60.6 & 11.0
 & \CCG 67.8 & \CCG \textbf{27.5} \\
\midrule
Comment removal & Incl. docstrings
 & 59.6 & 10.8
 & \underline{67.4} & \underline{19.5}
 & 65.1 & 17.6
 & 61.7 & 12.2
 & 55.1 & 5.9
 & \CCG \textbf{71.2} & \CCG \textbf{29.5} \\
\midrule
\multirow{3}{*}{Dead code} & 1 insertion
 & \textbf{78.8} & \textbf{28.0}
 & 52.6 & 13.3
 & 67.2 & 13.1
 & 63.7 & 12.0
 & 50.5 & 6.4
 & \CCG \underline{71.4} & \CCG \underline{25.1} \\
 & 3 insertions
 & \textbf{84.1} & \textbf{37.1}
 & 35.4 & 7.1
 & 66.0 & 12.5
 & 62.1 & 10.7
 & 38.5 & 3.0
 & \CCG \underline{68.4} & \CCG \underline{23.4} \\
 & 5 insertions
 & \textbf{85.5} & \textbf{42.7}
 & 29.5 & 6.3
 & 65.7 & 13.0
 & 61.2 & 9.8
 & 35.2 & 3.0
 & \CCG \underline{67.7} & \CCG \underline{24.7} \\
\midrule
\multirow{3}{*}{Control flow} & 1 step
 & 72.3 & 23.2
 & 72.0 & \underline{26.3}
 & \underline{72.5} & 26.0
 & 65.7 & 15.6
 & 61.1 & 11.0
 & \CCG \textbf{76.6} & \CCG \textbf{37.7} \\
 & 3 steps
 & \underline{72.3} & 23.3
 & 71.7 & 25.7
 & 71.8 & \underline{26.2}
 & 65.5 & 15.8
 & 60.4 & 10.7
 & \CCG \textbf{76.5} & \CCG \textbf{36.0} \\
 & 5 steps
 & \underline{72.4} & 23.0
 & 71.8 & 26.2
 & 72.1 & \underline{26.9}
 & 65.6 & 15.3
 & 60.0 & 10.5
 & \CCG \textbf{76.0} & \CCG \textbf{36.3} \\
\midrule
LLM rewriting & GPT-5 mini
 & \underline{55.1} & \textbf{8.4}
 & 48.8 & 6.4
 & \textbf{55.7} & 5.5
 & 49.9 & 4.5
 & 51.3 & 4.7
 & \CCG 51.4 & \CCG \underline{6.8} \\
\midrule
\multicolumn{2}{l}{AVG}
 & \textbf{73.2} & \underline{25.1}
 & 60.1 & 19.1
 & 68.3 & 19.4
 & 62.9 & 14.2
 & 54.6 & 8.7
 & \CCG \underline{70.1} & \CCG \textbf{27.2} \\
\bottomrule[1.2pt]
\end{tabular}
\end{table}

\paragraph{Dead-code insertion.}
Under dead-code insertion, \ours{} remains the second-best method in both
AUC and T@5 for all three strengths.
Unigram instead increases as more dead code is inserted, reaching
$85.5\%$ AUC and $42.7\%$ T@5 at five insertions.
Because newly inserted tokens can themselves alter the detector statistic,
this increase should be interpreted as behavior under the transformed
sequence rather than preservation of the original token-level evidence.

\paragraph{LLM rewriting.}
LLM rewriting is the most disruptive transformation for all methods.
For \ours{}, AUROC falls to $51.4\%$ and T@5 to $6.8\%$, while the other
methods also approach weak detection in this setting.
Unlike local edits, an LLM rewrite can simultaneously change token
identities, local contexts, and grammar states throughout the completion,
disrupting the keyed partitions reconstructed during replay.
Extensive rewriting therefore remains the most challenging transformation
for the current method.

\section{Conclusion}

We presented \ours{}, a grammar-guided code watermarking method 
that constructs keyed green-red partitions over grammar-admissible tokens 
and uses green temperature to strengthen the watermark according 
to the model's existing token preferences.
Across five models and five benchmarks spanning four programming languages,
\ours{} achieves a mean AUROC of $73.61\%$ and TPR@5\%FPR of $41.86\%$,
improving over the strongest corresponding baselines by $5.78$ and
$17.36$ percentage points, respectively.
Its mean Pass@1 is $59.18\%$, about $0.4$ points 
below the unwatermarked reference, 
showing that the stronger detection is achieved while largely
preserving functional correctness.
In particular, \ours{} attains the highest TPR@5\%FPR in all 25
model-benchmark configurations.
Ablations reveal complementary roles for the two components.
Grammar guidance improves mean Pass@1 by $2.23$ points 
with little change in aggregate AUROC, 
while green temperature improves mean AUROC by $9.05$ points 
and substantially strengthens low-FPR detection 
over additive bias with grammar retained.
These results show that grammar guidance 
primarily protects generation quality, 
while green temperature provides the stronger detectable signal.
\ours{} also retains competitive detection 
under a range of local code transformations, 
although LLM rewriting substantially weakens the signal.
Improving detection after large structural rewrites
and reducing the runtime overhead of the current implementation are 
important directions for future work.

\subsection*{AI use statement}

In this work, we used generative AI tools to assist with the implementation
of experimental code.
We have not used generative AI tools for other tasks requiring disclosure,
and the rest of the required-disclosure tasks are not applicable to this work.
Additionally, we used generative AI tools to edit the manuscript for
readability and language polishing.
We have reviewed all AI-assisted work.
AI-assisted experimental code was manually reviewed and tested for correctness,
and all AI-assisted manuscript edits were reviewed by the authors.
We take responsibility for the final content of this work,
including text, claims, or artifacts produced with the aid of generative AI.

\bibliography{iclr2027_conference}
\bibliographystyle{iclr2027_conference}

\clearpage
\appendix

\section{Detailed Algorithm}
\label{GTCW:app:sec:algorithm}

Algorithms~\ref{GTCW:app:alg:generation} and
\ref{GTCW:app:alg:detection} provide the detailed generation and replay
procedures for \ours{}.
The notation follows Section~\ref{GTCW:main:sec:gtcw}.

\paragraph{Partition implementation.}
\label{GTCW:app:par:partition_implementation}

The implementation derives the support signature $d_t$ by hashing a grammar
identifier together with the packed admissible-token mask.
A keyed hash then combines $d_t$ with the most recent $w$ token identifiers
to obtain the partition seed.
The seed induces a deterministic pseudo-random ranking of the admissible
tokens, of which $m_t$ are assigned to the green set according to
Eq.~\ref{GTCW:main:eq:state_seed}.
We use HMAC-SHA256 for the keyed hash and SplitMix64 for token ranking.
Generation and replay detection use the same deterministic procedure.

\subsection{Generation}
\label{GTCW:app:ssec:generation}

The entropy gate is evaluated before grammar masking and watermark
reweighting, so it reflects the original model uncertainty.
The grammar constraint remains active even when watermark insertion is
disabled.
For C++ and Java, the carrier gate $e_t$ becomes zero after 16 consecutive
comment tokens or four consecutive whitespace tokens in code.
The counters reset when the corresponding run ends and are reconstructed
during replay detection.
Non-finite model logits are excluded from entropy and sampling computations.
In all evaluated settings, logits of grammar-admissible source tokens are
finite, so $\mathcal B_t=\mathcal A_t$.

\begin{algorithm}[ht]
\caption{\ours{} source-token generation}
\label{GTCW:app:alg:generation}
\begin{algorithmic}[1]
\REQUIRE Prompt $x$, model $f_\theta$, grammar, key $k$,
and $\gamma,\tau,w,T_0,T_g$
\STATE Initialize the parser from the benchmark prefix and set $y\gets()$
\FOR{each source-token decoding step $t$}
\STATE $\ell_t\gets f_\theta(x,y_{<t})$
\STATE $\mathcal A_t\gets$ source-token IDs admitted by the parser
\STATE $\mathcal F_t\gets\{v\mid \ell_t(v)\text{ is finite}\}$
\STATE $\mathcal B_t\gets\mathcal A_t\cap\mathcal F_t$
\STATE $p_t(v)\gets
\exp(\ell_t(v))/\sum_{u\in\mathcal F_t}\exp(\ell_t(u))$
\STATE $H_t\gets-\sum_{v\in\mathcal F_t}p_t(v)\log p_t(v)$
\STATE Set $e_t$ from the comment and whitespace counters
\STATE $\widetilde\ell_t(v)\gets\ell_t(v)/T_0$
for $v\in\mathcal B_t$, and $-\infty$ otherwise
\IF{$e_t=1$ and $H_t>\tau$ and $|\mathcal A_t|>1$}
\STATE $c_t\gets$ the last $w$ token identifiers from the prompt and current prefix
\STATE Construct support signature $d_t$ from the current grammar state
\STATE $h_t\gets\mathcal H_k(d_t,c_t)$
\STATE Set $m_t$ by Eq.~\ref{GTCW:main:eq:state_seed}
and rank $\mathcal A_t$ using $h_t$
\STATE $\mathcal G_t\gets$ the first $m_t$ tokens in the ranking
\STATE $\mu_t\gets
\log\bigl(\sum_{v\in\mathcal B_t}\exp(\ell_t(v))/|\mathcal B_t|\bigr)$
\FOR{$v\in\mathcal G_t\cap\mathcal B_t$}
\STATE $\widetilde\ell_t(v)\gets
\mu_t/T_0+(\ell_t(v)-\mu_t)/T_g$
\ENDFOR
\ENDIF
\STATE Sample $y_t$ from $\operatorname{softmax}(\widetilde\ell_t)$ over
$\mathcal B_t$, append it to $y$, and consume it in the parser
\STATE Update the comment and whitespace counters
\ENDFOR
\RETURN $y$
\end{algorithmic}
\end{algorithm}

Algorithm~\ref{GTCW:app:alg:generation} describes the distribution
conditioned on generating a source token.
Special control tokens do not belong to the carrier set and are handled
separately.
Termination is allowed only when the parser state and completion policy
permit it.
When termination is available, the implementation applies a common offset
to source-token logits to preserve the unwatermarked grammar-constrained
probability of continuing versus stopping without changing the relative
probabilities among source tokens.
An empty finite admissible support cannot define a sampling distribution
and is treated as a generation failure.
Failed generations remain in the evaluation denominator.

\subsection{Detection by Replay}
\label{GTCW:app:ssec:detection}

Detection operates on the completed code, 
consisting of the original code prefix and the observed continuation.
The detector reconstructs the model context using the code prefix 
and the fixed prompting template, 
and obtains teacher-forced logits over the observed continuation.
At each step, it reconstructs the grammar state, entropy gate, 
and keyed green-red partition before consuming the observed token.
No generation-time trace or watermark metadata needs to be stored.
Admissible states with a single candidate are replayed normally 
but contribute no watermark evidence.
Replay terminates when the observed token is inadmissible, 
parsing fails, or no admissible token remains.
Evidence accumulated before termination is retained, 
so a finite score may be obtained from a partially replayed completion.

\begin{algorithm}[ht]
\caption{Replay-based detection on a candidate completion}
\label{GTCW:app:alg:detection}
\begin{algorithmic}[1]
\REQUIRE Code prefix $x$, continuation $y$, original model, grammar, key $k$,
and $\gamma,\tau,w$
\STATE Initialize the parser from the code prefix $x$
\STATE $S\gets0$, $V\gets0$
\FOR{$t=1,\ldots,|y|$}
\STATE Obtain $\mathcal A_t$ from the current parser state,
stopping on a parse error
\IF{$y_t\notin\mathcal A_t$}
\STATE Stop replay
\ENDIF
\IF{$|\mathcal A_t|>1$}
\STATE $\ell_t\gets f_\theta(x,y_{<t})$
\STATE $\mathcal F_t\gets\{v\mid\ell_t(v)\text{ is finite}\}$
\STATE $p_t(v)\gets
\exp(\ell_t(v))/\sum_{u\in\mathcal F_t}\exp(\ell_t(u))$
\STATE $H_t\gets-\sum_{v\in\mathcal F_t}p_t(v)\log p_t(v)$
\STATE Reconstruct $e_t$ from the comment and whitespace counters
\IF{$e_t=1$ and $H_t>\tau$}
\STATE $c_t\gets$ the last $w$ token identifiers from $x$ and $y_{<t}$
\STATE Construct support signature $d_t$ from the current grammar state
\STATE $h_t\gets\mathcal H_k(d_t,c_t)$
\STATE Set $m_t$ by Eq.~\ref{GTCW:main:eq:state_seed}
and rank $\mathcal A_t$ using $h_t$
\STATE $\mathcal G_t\gets$ the first $m_t$ tokens in the ranking
\STATE $\gamma_t\gets m_t/|\mathcal A_t|$
\STATE $S\gets S+\mathbf 1[y_t\in\mathcal G_t]-\gamma_t$
\STATE $V\gets V+\gamma_t(1-\gamma_t)$
\ENDIF
\ENDIF
\STATE Consume $y_t$ in the parser, stopping if consumption fails
\STATE Update the comment and whitespace counters
\ENDFOR
\RETURN $S/\sqrt V$ if $V>0$, and no score otherwise
\end{algorithmic}
\end{algorithm}

\section{Experimental Details}
\label{GTCW:app:sec:experimental_details}

\subsection{Implementation and Decoding Settings}
\label{GTCW:app:ssec:decoding_settings}

For each task and seed, all methods generate one sampled completion 
using the same decoding configuration. 
We use temperature $T_0=1.0$, top-$p=1.0$, no top-$k$ filtering, 
a repetition penalty of $1.0$, and a maximum of 512 new tokens. 
Models are loaded in bfloat16.
The watermark is applied only to the generated continuation, 
while preceding tokens remain available 
to context-dependent watermarking methods.
For \ours{}, we use $\gamma=0.5$, 
entropy threshold $\tau=0.9$ nats,
green temperature $T_g=0.5$, and context width $w=1$.
Experiments use seeds 42, 43, and 44.

\paragraph{Baseline implementations.}
We integrate Unigram, KGW, SWEET, STONE, and STA-1 
into the same evaluation 
pipeline~\citep{ZhaoALW2024,KirchenbauerGWKMG2023,LeeHAHLYSK2024,KimPH2026,MaoWCFC2025}.
Unigram uses a fixed keyed partition, 
while KGW uses a context-dependent partition. 
SWEET additionally applies an entropy gate with threshold $0.9$.
All baselines use $\gamma=0.5$.
The additive-bias methods Unigram, KGW, SWEET, and STONE use $\delta=2.0$,
while STA-1 does not use an additive-bias parameter.
These settings are fixed across models and benchmarks.
Only \ours{} applies grammar constraints in the main comparison.
The ablations separately remove grammar guidance 
and replace green temperature with additive bias.

\paragraph{STONE implementation and settings.}
For a fair comparison, we make two corrections 
to the publicly released STONE implementation.
First, we correct the scored-token count used by the detector 
so that its normalization reflects the positions selected for watermarking.
Second, we set \texttt{prefix\_length=1}, 
which makes the green-red partition depend on the preceding token 
rather than remaining fixed across decoding steps.
We use $\gamma=0.5$ and $\delta=2.0$, 
consistent with the other additive green-list baselines.
All reported STONE results use this corrected configuration 
for both generation and detection.

\subsection{Datasets, Prompts, and Metrics}
\label{GTCW:app:ssec:dataset_metrics}

\paragraph{Benchmarks and test suites.}
We evaluate HumanEval+ and MBPP+ through EvalPlus using 164 and
378 tasks, respectively~\citep{LiuXWZ2023}.
Functional correctness is measured on the full augmented test suites.
For C++, Java, and Go, we use the 164-task HumanEvalPack subset 
for each language together 
with its execution tests~\citep{MuennighoffLZZHZSTWL2024}.

\paragraph{Continuation prompts.}
All benchmarks are evaluated as code continuation tasks.
Each model receives an unfinished code prefix 
through the following template.
\begin{quote}\small\ttfamily\raggedright
Complete the following <language> code. Return only the completed code without explanations.\par
\medskip
<benchmark prefix>
\end{quote}
We apply the checkpoint-specific chat template 
and prefill the assistant response with the same benchmark prefix 
so that generation continues the code directly.
The prefix, including trailing whitespace, is preserved exactly.
Generated presentation wrappers such as code fences are removed 
before evaluation, 
after which the language-specific pipeline reconstructs the executable program.
For replay detection, the original code prefix is retained 
and the fixed model prompt is reconstructed from the same template.

\paragraph{Negative class.}
The negative class consists of unwatermarked completions generated with
the same model, task, decoding configuration, 
and seed as the corresponding watermarked outputs.
This controls for differences in model family, prompting, and task content
when measuring watermark evidence.
Accordingly, the reported detection results characterize separation 
from matched unwatermarked model outputs rather than human-written code.

\paragraph{AUROC and score coverage.}
For one model-benchmark-seed setting, let $z_i^+$ and $z_j^-$ denote the
positive and negative detection scores, with class sizes $N_+$ and $N_-$.
Missing or non-finite scores in either class receive a common value below
all finite scores.
AUROC is computed as
\begin{equation}
A=\frac{1}{N_+N_-}
\sum_{i=1}^{N_+}\sum_{j=1}^{N_-}
\left(
\mathbf 1[z_i^+>z_j^-]
+\frac12\mathbf 1[z_i^+=z_j^-]
\right).
\label{GTCW:app:eq:auroc}
\end{equation}
Let $b_i$ indicate whether the original score is finite.
Finite-score coverage is $C=N^{-1}\sum_i b_i$.
For \ours{}, missing scores can arise 
when no eligible watermark position is available 
or when grammar replay terminates before watermark evidence is obtained.

\paragraph{TPR@5\%FPR.}
For threshold $u$, an output is detected 
only if its score is finite and at least $u$.
Writing the corresponding decisions as $D_i^+(u)$ and $D_j^-(u)$,
\begin{equation}
\operatorname{TPR}(u)
=
\frac{\sum_i D_i^+(u)}{N_+},
\qquad
\operatorname{FPR}(u)
=
\frac{\sum_j D_j^-(u)}{N_-}.
\end{equation}
Let $\mathcal U$ contain the distinct finite negative scores together 
with a value immediately above their maximum.
For target false positive rate $\alpha$, we use
\begin{equation}
u_\alpha
=
\min\{u\in\mathcal U
\mid \operatorname{FPR}(u)\leq\alpha\},
\qquad
\operatorname{TPR}@\alpha
=
\operatorname{TPR}(u_\alpha).
\label{GTCW:app:eq:tpr}
\end{equation}
We use $\alpha=0.05$ throughout the reported TPR@5\%FPR results.
Unscorable outputs remain in the class denominators and are never detected.
Because thresholds are selected from observed negative scores 
without interpolation, the achieved FPR can be below $5\%$.

\paragraph{Functional correctness and joint detection.}
Let $P_i$ indicate whether completion $i$ passes all required execution tests.
With one completion per task and seed, Pass@1 is
$N^{-1}\sum_i P_i$.
Using the same threshold $u_\alpha$, we additionally measure detection among
correct outputs and joint correctness and detection,
\begin{equation}
Q_\alpha
=
\frac{\sum_i P_iD_i^+(u_\alpha)}
{\sum_i P_i},
\qquad
J_\alpha
=
\frac{1}{N_+}
\sum_i P_iD_i^+(u_\alpha).
\label{GTCW:app:eq:joint_detection}
\end{equation}
Here $Q_\alpha$ is the detection rate conditioned on functional correctness,
while $J_\alpha$ is the fraction of all watermarked outputs 
that are both correct and detected.
The threshold is calibrated on the full negative set and is not recomputed
for the correct-output subset.
Failed generations and test failures remain in the corresponding denominators.

\paragraph{Aggregation.}
Metrics are first computed for each model, benchmark, and seed.
Table entries average the three seeds, 
and overall results weight model-benchmark settings equally 
so that benchmarks with more tasks do not receive greater weight.
Transformation results similarly average the five benchmarks and three seeds on
Qwen2.5-Coder~(3B), with the reported AVG giving equal weight to the 13
transformation settings.

\subsection{Models}
\label{GTCW:app:ssec:models}

We evaluate five instruction-tuned checkpoints from several model families.
Qwen2.5-Coder 3B and 7B provide a code-specialized pair 
at two model scales,\footnote{\texttt{Qwen/Qwen2.5-Coder-3B-Instruct} 
and \texttt{Qwen/Qwen2.5-Coder-7B-Instruct}.}
allowing us to examine whether the watermark behaves 
consistently within the same family.
Qwen3.5 9B\footnote{\texttt{Qwen/Qwen3.5-9B}.}
extends the evaluation beyond the Qwen2.5-Coder architecture,
while Llama3.1 8B\footnote{\texttt{meta-llama/Llama-3.1-8B-Instruct}.}
and Gemma4 12B\footnote{\texttt{google/gemma-4-12B-it}.}
broaden the comparison across distinct model families and tokenizers.
Gemma4 is also the largest checkpoint in our evaluation.
We abbreviate Qwen2.5-Coder as Qwen2.5-C in tables.
Sensitivity, ablation, and transformation experiments 
use Qwen2.5-Coder (3B) as a fixed source model.

\subsection{Grammars}
\label{GTCW:app:ssec:grammars}

We use grammars for Python, C++, Java, and Go together with an independent
Rust grammar engine for admissible-token computation.
The Python grammar is adapted from the Lark Python grammar 
used by SynCode~\citep{UgareSKMS2025}, 
with string literals separated from comments.
The Java and Go grammars are similarly 
adapted to the benchmark continuation setting.
For C++, we use a task-oriented grammar designed for function completion.
It permits some constructions beyond a strict language specification, 
so grammar admissibility guarantees only compatibility with the grammar 
and does not imply successful compilation or functional correctness.
Execution tests are used separately to evaluate these properties.
Since benchmark prefixes are not always complete translation units,
parser-side adapters provide the surrounding context 
required to initialize the grammar state.
C++ and Java use synthetic function-body wrappers, 
while Go may use a virtual package prefix.
These wrappers affect only parser initialization 
and are not added to the code presented to the model.
Grammar definitions, lexer information, token bytes, and admissibility data
are precompiled into artifacts consumed by the Rust runtime.
Generation and replay detection use the same grammar artifacts and parser
implementation.

\subsection{Code Transformation Evaluation Protocol}
\label{GTCW:app:ssec:transformation_protocol}

We evaluate six transformation families with 13 settings on Qwen2.5-Coder~(3B)
outputs from the five benchmarks and seeds 42, 43, and 44.
Each transformation is applied independently 
to the corresponding clean completion, 
and only the generated continuation is modified.
The original code prefix is preserved for reconstruction and replay detection.
All watermarking methods are evaluated with the same transformation
implementations and strength settings.
Local edit selection uses deterministic per-record seeds 
derived from the record identity, method, transformation, and setting.
The LLM rewrite is stochastic but uses the same request procedure
across all methods.
Matched clean Base completions serve as the negative class for all
transformation settings.
Consequently, the reported metrics characterize detection after transformation
relative to the clean negative distribution and may also reflect
transformation-induced score shifts.
The reported AVG averages the 13 transformation settings and excludes
the clean reference.

\paragraph{Identifier renaming.}
We rename 25\%, 50\%, 75\%, or 100\% of eligible local variables.
Tree-sitter is used to identify declarations and references 
within the generated continuation.
We exclude ambiguous bindings, references that cross the code-prefix boundary
or nested functions, and member or API identifiers.
Selected variables and their accepted references are assigned fresh names
that do not collide with existing identifiers or language keywords.
The requested rate is measured over eligible variables.

\paragraph{Formatting and comment removal.}
Formatting uses Black for Python, clang-format for C++,
google-java-format for Java, and gofmt for Go.
A temporary boundary marker is used to separate 
the preserved code prefix from the formatted continuation.
Comment removal deletes removable comments while preserving line structure.
For Python, docstrings in the generated continuation are also removed, 
with \texttt{pass} inserted when required to keep a body nonempty.
Comments in the code prefix and protected directives 
such as Go build directives are left unchanged.

\paragraph{Dead-code insertion.}
We request 1, 3, or 5 insertions at eligible statement boundaries 
inside functions.
Each insertion adds a constant-false branch 
containing a local dummy assignment.
For Go, the inserted block also includes a blank-identifier 
use of the dummy variable.
Insertion sites are selected in deterministic pseudorandom order.
When fewer eligible sites are available, 
all available sites are used and the achieved insertion count is recorded.

\paragraph{Control-flow transformations.}
We request 1, 3, or 5 nonoverlapping edits at eligible syntax nodes.
The implemented transformations include double negation of conditions,
condition inversion with branch exchange, 
expansion of an else-if chain into nested branches, 
and equivalent Boolean-literal rewrites.
Each requested step corresponds to one selected edit site.
Overlapping or unsupported sites are excluded, 
and the achieved number of edits is recorded 
when fewer sites are available than requested.

\paragraph{LLM rewriting.}
We use \texttt{gpt-5-mini} to generate one rewritten continuation 
through a structured JSON response.
The request includes the language, preserved code prefix, 
and original completion, 
and asks for changes to control flow, data flow, or implementation
structure while preserving the public API and program behavior.
The response is restricted to code 
and may not introduce additional tests or Markdown wrappers.
For MBPP+, where the generated continuation may include the function declaration, 
the public function name and signature are preserved.
Invalid responses, unresolved prefix echoes, unchanged rewrites, 
and request failures are retried up to three times.
If all attempts fail, the original completion is retained.
The rewriter receives neither the watermark key nor detector feedback.

\paragraph{Validation and failure handling.}
After transformation, the preserved code prefix 
and transformed continuation must pass Tree-sitter parsing.
Transformations that fail this check fall back to the original completion.
Successful partial transformations are retained, 
and both requested and achieved strengths are recorded.
All outputs, including unchanged and failed transformations, 
remain in the reported metrics.
These transformations are designed to preserve program behavior,
but functional tests are not rerun after transformation.

\paragraph{Detection and aggregation.}
Watermark scores are recomputed on the transformed continuation 
using the original code prefix and the corresponding method configuration.
For SWEET, model entropy is recomputed on the transformed token sequence
using the generation-time entropy threshold.
STONE uses its corrected detector with
\texttt{prefix\_length=1} and $\delta=2.0$.
AUROC and TPR@5\%FPR follow
Appendix~\ref{GTCW:app:ssec:dataset_metrics}.
Results are averaged equally over the five benchmarks and three seeds,
and AVG gives equal weight to the 13 transformation settings while excluding
the clean reference.

\subsection{Experimental Environment}
\label{GTCW:app:ssec:environment}

Experiments were run on Rocky Linux 9.6 with an
AMD Ryzen Threadripper 9960X 24-core CPU, 250 GiB of system memory,
and NVIDIA RTX PRO 6000 Blackwell Max-Q GPUs with approximately
96 GiB of memory each.
Each model process uses a single GPU.
The system uses NVIDIA driver 580.105.08 and CUDA 13.0.
The software environment uses Python 3.12.13,
PyTorch 2.13.0 with CUDA 13.0,
Transformers 5.14.1, and SynCode 0.4.16.
The Rust grammar backend is version 0.2.0 and is built with Rust 1.98.0
using the Rust 2021 edition.
Models are loaded in bfloat16.
For code formatting, we use Black 26.5.1 for Python,
clang-format 23.1.0 for C++,
google-java-format 1.28.0 for Java,
and gofmt from Go 1.26.5 for Go.
The execution environment uses GCC 11.5.0 for C++ and
OpenJDK 17.0.18 for Java.
Tree-sitter 0.25.2 is used with language bindings
0.25.0 for Python, 0.23.4 for C++, 0.23.5 for Java,
and 0.25.0 for Go.

\section{Additional Experimental Results}
\label{GTCW:app:sec:additional_experiments}

This section provides additional analyses 
that complement the main results from several perspectives.
We first report the complete model-benchmark breakdown, 
including results on the additional recent open-weight models 
Qwen3.5~(9B) and Gemma4~(12B), and
show that the main detection trends remain consistent beyond the models
highlighted in the main text.
We then examine detection specifically on functionally correct outputs 
using the conditional and joint metrics defined in
Appendix~\ref{GTCW:app:ssec:dataset_metrics}, 
followed by wall-clock runtime measurements of the current implementation.
Finally, qualitative code examples illustrate 
how different watermarking methods distribute their signal 
across generated code and how \ours{} combines grammar admissibility, 
carrier selection, and green-temperature reweighting at the token level.

\subsection{Full Main Results}
\label{GTCW:app:ssec:full_main_results}

Table~\ref{GTCW:app:tab:full_main_results} extends the main evaluation
with two additional recent open-weight models, 
Qwen3.5~(9B) and Gemma4~(12B). 
The main trend observed in the models reported in the main text 
remains visible in these additional results, 
particularly at the low-FPR operating point.
On Qwen3.5, \ours{} achieves the highest AUC and T@5 
on all five benchmarks.
The T@5 gains are especially pronounced on C++ and Go, 
where \ours{} reaches 47.8 and 52.2, compared with 31.7 and 22.2 
for the strongest baseline, respectively.
On Gemma4, the AUC advantage is less uniform, 
but \ours{} still achieves the highest T@5 on all five benchmarks.
This distinction is most visible on C++, Java, and Go, 
where AUROC is competitive rather than uniformly best 
while low-FPR detection remains strongest.

The correctness results also show that 
the detection gains do not follow a single quality pattern across models.
Qwen3.5 exhibits some P@1 degradation on the HumanEvalPack languages,
whereas Gemma4 remains close to Base on C++ and Java 
and does not show a consistent loss across the other benchmarks.
Together, these results extend the detectability-correctness behavior
observed in the main text to two additional recent open-weight models and
show that the most consistent advantage of \ours{} is 
its detection strength at low false-positive rates.

\begin{table}[t]
\centering
\caption{
Full detection and functional-correctness results 
across five models and five benchmarks.
AUC denotes AUROC, T@5 denotes TPR@5\%FPR, and P@1 denotes Pass@1.
Values are mean percentages over three seeds.
Gray shading indicates \ours{}.
Bold and underline mark the best and second-best watermarked methods,
respectively, based on unrounded values.
}
\label{GTCW:app:tab:full_main_results}
\small
\setlength{\tabcolsep}{1.5pt}
\begin{tabular}{@{}cc*{5}{rrr}@{}}
\toprule[1.2pt]
& & \multicolumn{3}{c}{HumanEval+} & \multicolumn{3}{c}{MBPP+} & \multicolumn{9}{c}{HumanEvalPack} \\
\cmidrule(lr){3-5}\cmidrule(lr){6-8}\cmidrule(lr){9-17}
Model & Method & \multicolumn{3}{c}{Python} & \multicolumn{3}{c}{Python} & \multicolumn{3}{c}{C++} & \multicolumn{3}{c}{Java} & \multicolumn{3}{c}{Go} \\
\cmidrule(lr){3-5}\cmidrule(lr){6-8}\cmidrule(lr){9-11}\cmidrule(lr){12-14}\cmidrule(lr){15-17}
& & AUC & T@5 & P@1 & AUC & T@5 & P@1 & AUC & T@5 & P@1 & AUC & T@5 & P@1 & AUC & T@5 & P@1 \\
\midrule[1.2pt]
\multirow{7}{*}{\shortstack{Qwen2.5-C\\(3B)}} & Base & --- & --- & 67.9 & --- & --- & 52.1 & --- & --- & 52.2 & --- & --- & 71.8 & --- & --- & 51.2 \\
 & Unigram & 68.1 & 18.9 & 65.0 & 82.4 & 38.5 & 43.5 & 75.5 & 28.5 & 46.8 & 67.2 & 16.9 & 63.4 & 71.2 & 15.5 & 45.7 \\
 & KGW & 68.1 & 22.0 & 63.4 & 81.2 & 39.7 & 47.6 & \underline{76.7} & \underline{33.9} & 45.3 & 65.5 & 14.4 & 66.9 & 68.6 & 22.0 & 47.6 \\
 & SWEET & \underline{68.9} & \underline{24.4} & 68.7 & \underline{84.4} & \underline{41.8} & 50.8 & 71.0 & 30.1 & \textbf{53.9} & \underline{68.5} & \underline{25.8} & \textbf{71.1} & \underline{72.8} & \underline{29.3} & 48.2 \\
 & STONE & 63.3 & 11.2 & 67.1 & 74.3 & 24.1 & 47.1 & 66.4 & 16.3 & 49.2 & 61.8 & 11.4 & \underline{70.3} & 63.7 & 16.1 & 46.5 \\
 & STA-1 & 57.9 & 7.9 & \textbf{70.5} & 70.5 & 20.0 & \underline{51.2} & 62.8 & 11.2 & \underline{51.0} & 59.8 & 11.4 & 69.3 & 60.5 & 8.9 & \textbf{51.8} \\
 & \CCG \ours{} & \CCG \textbf{77.6} & \CCG \textbf{49.2} & \CCG \underline{69.5} & \CCG \textbf{96.5} & \CCG \textbf{83.6} & \CCG \textbf{51.8} & \CCG \textbf{81.1} & \CCG \textbf{51.4} & \CCG 50.0 & \CCG \textbf{75.9} & \CCG \textbf{37.2} & \CCG \underline{70.3} & \CCG \textbf{86.4} & \CCG \textbf{56.5} & \CCG \underline{49.2} \\
\midrule
\multirow{7}{*}{\shortstack{Qwen2.5-C\\(7B)}} & Base & --- & --- & 77.0 & --- & --- & 61.3 & --- & --- & 63.6 & --- & --- & 79.3 & --- & --- & 63.2 \\
 & Unigram & 60.5 & \underline{12.2} & 74.4 & \underline{80.6} & 32.6 & 53.1 & \underline{67.3} & \underline{17.3} & 60.4 & \underline{59.2} & 9.4 & \underline{78.9} & \textbf{62.7} & 10.2 & 61.8 \\
 & KGW & \underline{60.6} & 7.9 & 75.2 & 79.4 & \underline{40.0} & 55.0 & \textbf{67.9} & 15.5 & 56.9 & \textbf{60.9} & 8.7 & 74.6 & 60.6 & \underline{11.8} & 59.4 \\
 & SWEET & 57.0 & 3.9 & 76.8 & 75.4 & 27.3 & \underline{60.0} & 59.1 & 11.6 & \underline{62.6} & 53.8 & \underline{11.2} & 77.2 & 57.7 & 8.3 & \textbf{63.0} \\
 & STONE & 59.3 & 10.8 & 76.6 & 69.8 & 20.9 & 57.6 & 58.7 & 7.5 & 62.0 & 55.7 & 6.7 & 77.6 & 58.2 & 10.8 & \textbf{63.0} \\
 & STA-1 & 57.3 & 10.0 & \textbf{77.2} & 65.6 & 15.8 & \textbf{60.9} & 58.6 & 8.1 & \textbf{63.2} & 55.1 & 7.9 & 76.0 & 55.6 & 7.3 & 60.2 \\
 & \CCG \ours{} & \CCG \textbf{60.9} & \CCG \textbf{21.5} & \CCG \underline{77.0} & \CCG \textbf{88.3} & \CCG \textbf{56.0} & \CCG 59.5 & \CCG 64.8 & \CCG \textbf{27.4} & \CCG 58.1 & \CCG 56.7 & \CCG \textbf{16.1} & \CCG \textbf{80.1} & \CCG \underline{62.6} & \CCG \textbf{20.9} & \CCG \underline{62.2} \\
\midrule
\multirow{7}{*}{\shortstack{Qwen3.5\\(9B)}} & Base & --- & --- & 73.0 & --- & --- & 66.8 & --- & --- & 66.9 & --- & --- & 79.9 & --- & --- & 63.8 \\
 & Unigram & 63.9 & 14.6 & 68.5 & 61.8 & 12.8 & 64.1 & 71.4 & 22.4 & 60.0 & \underline{66.7} & 15.2 & 71.8 & 66.7 & 16.7 & 53.9 \\
 & KGW & \underline{67.5} & \underline{19.1} & 72.4 & \underline{66.0} & \underline{18.8} & 63.2 & \underline{74.0} & \underline{31.7} & 57.1 & 62.8 & 14.0 & 72.0 & 67.9 & 16.5 & 55.1 \\
 & SWEET & 64.2 & 15.2 & 73.0 & 62.8 & 18.3 & 66.1 & 68.7 & 29.5 & \underline{65.5} & 64.7 & \underline{27.0} & \textbf{79.3} & \underline{70.1} & \underline{22.2} & \textbf{63.2} \\
 & STONE & 62.5 & 11.8 & 70.3 & 60.5 & 12.8 & 64.0 & 66.1 & 21.1 & 62.4 & 59.1 & 12.2 & 74.6 & 64.3 & 16.1 & 59.8 \\
 & STA-1 & 57.3 & 7.9 & \textbf{74.8} & 57.5 & 7.9 & \textbf{67.6} & 59.0 & 11.8 & \textbf{66.9} & 59.8 & 9.8 & 75.8 & 57.8 & 9.8 & \underline{60.6} \\
 & \CCG \ours{} & \CCG \textbf{72.2} & \CCG \textbf{36.4} & \CCG \underline{73.8} & \CCG \textbf{69.1} & \CCG \textbf{33.4} & \CCG \underline{66.5} & \CCG \textbf{75.5} & \CCG \textbf{47.8} & \CCG 63.0 & \CCG \textbf{69.2} & \CCG \textbf{31.3} & \CCG \underline{76.8} & \CCG \textbf{82.7} & \CCG \textbf{52.2} & \CCG 56.1 \\
\midrule
\multirow{7}{*}{\shortstack{Llama3.1\\(8B)}} & Base & --- & --- & 50.6 & --- & --- & 57.8 & --- & --- & 40.0 & --- & --- & 53.1 & --- & --- & 32.9 \\
 & Unigram & \underline{70.2} & 21.5 & 46.8 & \underline{63.4} & 12.3 & 56.4 & 76.7 & 24.6 & 34.2 & 70.2 & 22.4 & 45.5 & 71.0 & 26.0 & 23.8 \\
 & KGW & 66.6 & 23.0 & 48.4 & 63.0 & \underline{15.5} & 55.8 & 76.7 & 31.3 & 33.5 & 70.3 & 25.2 & 47.6 & 73.9 & 25.6 & 30.9 \\
 & SWEET & 69.9 & \underline{27.2} & 49.4 & 62.1 & 15.1 & 57.0 & \underline{81.2} & \underline{41.5} & \underline{36.8} & \underline{73.3} & \underline{33.3} & 51.6 & \underline{82.6} & \underline{47.0} & 30.5 \\
 & STONE & 65.5 & 16.9 & 48.0 & 60.6 & 12.9 & 56.4 & 69.6 & 21.5 & \underline{36.8} & 63.4 & 13.0 & 51.4 & 69.0 & 22.0 & 29.5 \\
 & STA-1 & 55.7 & 10.4 & \underline{51.4} & 56.0 & 9.0 & \underline{57.6} & 64.8 & 14.6 & \textbf{39.2} & 59.1 & 17.1 & \underline{52.0} & 62.1 & 13.4 & \textbf{36.0} \\
 & \CCG \ours{} & \CCG \textbf{78.6} & \CCG \textbf{45.9} & \CCG \textbf{53.1} & \CCG \textbf{66.6} & \CCG \textbf{27.7} & \CCG \textbf{58.6} & \CCG \textbf{90.5} & \CCG \textbf{76.2} & \CCG 33.7 & \CCG \textbf{79.6} & \CCG \textbf{47.0} & \CCG \textbf{52.4} & \CCG \textbf{92.8} & \CCG \textbf{73.0} & \CCG \underline{31.5} \\
\midrule
\multirow{7}{*}{\shortstack{Gemma4\\(12B)}} & Base & --- & --- & 52.0 & --- & --- & 61.6 & --- & --- & 53.5 & --- & --- & 52.6 & --- & --- & 45.5 \\
 & Unigram & \underline{66.4} & 22.8 & \underline{52.9} & 59.3 & 13.6 & 59.2 & 61.1 & 15.7 & 50.2 & 58.2 & 13.8 & 50.2 & \underline{61.7} & 12.0 & 44.9 \\
 & KGW & 65.8 & 18.3 & 51.2 & \textbf{62.2} & 13.8 & 60.4 & \textbf{64.6} & 17.1 & \underline{53.5} & \textbf{62.8} & 15.0 & 47.4 & \textbf{62.2} & 15.2 & 44.7 \\
 & SWEET & 65.2 & \underline{32.5} & 51.8 & 57.8 & \underline{25.6} & 61.6 & 58.7 & \underline{24.4} & \textbf{54.5} & 56.9 & \underline{21.5} & \textbf{52.6} & 55.5 & \underline{18.5} & \underline{46.8} \\
 & STONE & 60.1 & 16.9 & 52.0 & 59.5 & 15.4 & 60.8 & 59.8 & 13.0 & 52.2 & 59.2 & 12.6 & 45.5 & 57.2 & 10.0 & 44.9 \\
 & STA-1 & 59.8 & 11.8 & 51.2 & 58.0 & 10.5 & \underline{61.7} & 58.6 & 10.4 & 52.9 & 55.6 & 9.8 & \underline{52.2} & 58.7 & 8.3 & 45.7 \\
 & \CCG \ours{} & \CCG \textbf{70.7} & \CCG \textbf{47.2} & \CCG \textbf{70.9} & \CCG \underline{60.1} & \CCG \textbf{29.4} & \CCG \textbf{62.7} & \CCG \underline{62.8} & \CCG \textbf{30.5} & \CCG 52.9 & \CCG \underline{61.7} & \CCG \textbf{29.1} & \CCG 51.6 & \CCG 57.3 & \CCG \textbf{19.7} & \CCG \textbf{48.0} \\
\bottomrule[1.2pt]
\end{tabular}%
\end{table}

\subsection{Detection on Functionally Correct Outputs}
\label{GTCW:app:ssec:correct_output_detection}

For code watermarking, detection on functionally correct outputs is
particularly important 
because only these completions constitute usable solutions.
We therefore complement the overall detection results 
with the conditional and joint metrics defined in
Appendix~\ref{GTCW:app:ssec:dataset_metrics}.
$Q$ measures the fraction of functionally correct 
watermarked outputs that are detected, 
while $J$ measures the fraction of all watermarked outputs
that are both functionally correct and detected.

Table~\ref{GTCW:app:tab:joint_detection} shows that 
\ours{} achieves the strongest result under both measures.
Among functionally correct outputs, \ours{} detects $28.01\%$, 
compared with $14.50\%$ for the strongest baseline, SWEET.
Its joint success rate is $15.80\%$, nearly twice SWEET's $8.16\%$.
Thus, \ours{} places detectable watermark evidence 
in a substantially larger fraction of outputs 
that also satisfy the execution tests.

The conditional detection rate of \ours{} is lower than 
its overall TPR@5\%FPR of $41.86\%$, 
and the same trend appears for all baselines.
This shows that detection measured over all generated outputs does not
fully reflect watermark retention in functionally correct code.
The gap is especially relevant for code generation, 
where detection on incorrect completions contributes little 
to provenance of usable programs.

\ours{} has finite-score coverage of $74.07\%$, 
similar to SWEET at $74.42\%$.
Both methods can produce unscorable outputs 
when no entropy-eligible position is available, 
while \ours{} additionally relies on successful grammar replay.
Unscorable outputs remain in the denominators of $Q$ and $J$ as defined
in Appendix~\ref{GTCW:app:ssec:dataset_metrics}.

\begin{table}[ht]
\centering
\caption{
Detection on functionally correct outputs at TPR@5\%FPR.
$Q$ denotes detection among correct outputs, 
and $J$ denotes outputs that are both correct and detected.
Coverage is the fraction with finite scores.
Values are percentages averaged over five models, five benchmarks, and three seeds.
}
\label{GTCW:app:tab:joint_detection}
\small
\setlength{\tabcolsep}{8pt}
\begin{tabular}{lrrrrr}
\toprule[1.2pt]
\multirow{2}{*}{Method} & \multirow{2}{*}{Pass@1} & Detected among & Correct and & \multirow{2}{*}{Coverage} & \multirow{2}{*}{FPR} \\
 & & correct~($Q$) & detected~($J$) & & \\
\midrule[1.2pt]
Base & 59.58 & --- & --- & --- & --- \\
Unigram & 55.00 & 10.12 & 5.29 & 100.00 & 4.84 \\
KGW & 55.40 & 12.51 & 6.70 & 99.97 & 4.85 \\
SWEET & 58.87 & 14.50 & 8.16 & 74.42 & 3.74 \\
STONE & 57.03 & 9.84 & 5.38 & 99.90 & 4.83 \\
STA-1 & 59.08 & 7.73 & 4.58 & 100.00 & 4.84 \\
\CCG \ours{} & \CCG 59.18 & \CCG 28.01 & \CCG 15.80 & \CCG 74.07 & \CCG 4.46 \\
\bottomrule[1.2pt]
\end{tabular}
\end{table}

\subsection{Generation Wall-clock Time}
\label{GTCW:app:ssec:runtime}

We additionally measure wall-clock generation time and token throughput.
Timing includes model generation and the online watermarking procedure,
with device synchronization, but excludes model loading, offline grammar
precompilation, replay detection, and functional evaluation.
For each model-benchmark-seed setting, 
we first average completion time, generated token count, 
and per-completion token throughput, and then
average the resulting values equally across settings.
Token counts refer to raw generated tokens before cleanup.

As shown in Table~\ref{GTCW:app:tab:generation_time}, 
\ours{} requires $11.50$ seconds per completion, 
compared with $4.28$ seconds for Base
and $4.40$--$4.62$ seconds for the other watermarking methods.
Its mean throughput is $17.24$ tokens per second, 
compared with $63.07$--$64.93$ for the other methods.
Although the grammar backend is implemented in Rust 
and grammar artifacts are precompiled, 
the current \ours{} implementation is not yet optimized
for runtime and introduces substantial online overhead.
We regard this as an implementation limitation of the current system.
Further optimization of the integration between model decoding and
grammar-state processing may reduce this cost.

\begin{table}[ht]
\centering
\caption{
Generation wall-clock time, output length, and throughput.
Values are averaged equally over five models, five benchmarks, and three seeds.
Throughput is averaged per completion.
}
\label{GTCW:app:tab:generation_time}
\small
\setlength{\tabcolsep}{12pt}
\begin{tabular}{lrrr}
\toprule[1.2pt]
Method & Seconds/completion & Tokens/completion & Tokens/second \\
\midrule[1.2pt]
Base & 4.28 & 175.85 & 64.88 \\
Unigram & 4.62 & 192.68 & 64.93 \\
KGW & 4.56 & 187.60 & 63.70 \\
SWEET & 4.41 & 179.10 & 63.37 \\
STONE & 4.46 & 180.10 & 63.07 \\
STA-1 & 4.40 & 178.45 & 63.49 \\
\CCG \ours{} & \CCG 11.50 & \CCG 158.01 & \CCG 17.24 \\
\bottomrule[1.2pt]
\end{tabular}
\end{table}

\subsection{Generated Code Examples}
\label{GTCW:app:ssec:code_examples}

\begin{figure}[ht]
\centering
\includegraphics[width=\linewidth]{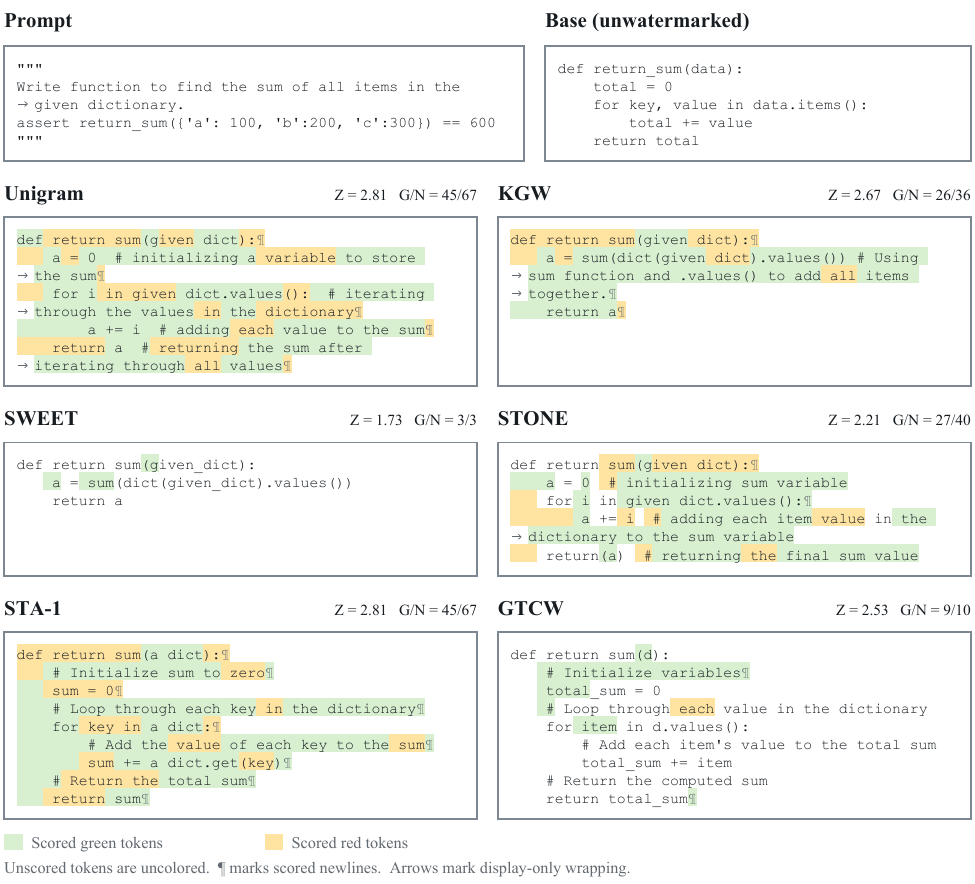}
\caption{
Functionally correct and detected outputs for MBPP+/796 using
Qwen2.5-Coder~(3B).
All displayed completions pass the augmented tests and exceed their
corresponding TPR@5\%FPR detection thresholds.
$Z$ denotes the detector score and $G/N$ the number of green tokens 
among scored positions.
Green highlighting marks scored green tokens.
}
\label{GTCW:app:fig:code_examples}
\end{figure}

Figure~\ref{GTCW:app:fig:code_examples} shows
functionally correct and detected outputs for MBPP+/796,
which asks for the sum of all values in a dictionary.
All methods use Qwen2.5-Coder~(3B) with the same code prefix,
and every displayed completion passes the full MBPP+ test suite.

The examples highlight how the methods differ in 
where watermark evidence is introduced.
Unigram, KGW, and STA-1 apply their watermarking rule broadly 
across generated positions, including identifiers, syntax, and comments.
For example, in the STA-1 output, repeated occurrences of the local
identifier \texttt{sum} are independently watermarkable.
Such broad carrier selection provides many opportunities 
to accumulate signal, 
but also perturbs token choices at positions that may be important
for the generated program.
STONE excludes some syntax-critical token categories, 
but in this example a substantial portion of its scored signal appears in comments.
SWEET instead uses entropy gating to avoid low-entropy positions.
Its concise completion happens to contain only three scored positions,
all of which are green.

\ours{} uses a more restrictive carrier condition 
that combines grammar admissibility, the carrier gate $e_t$, 
and the entropy gate.
Its advantage therefore does not come from watermarking 
a broader set of positions.
Rather, green temperature strengthens the signal 
within the selected admissible positions 
while preserving the model's relative preference among green candidates.
In the displayed completion, ten positions are scored and nine are green.
The remaining scored red token further illustrates that \ours{} biases,
rather than forces, the green choice.
Overall, the example reflects the intended design of \ours{}:
strong watermark evidence from selectively chosen positions without
requiring broad token-level perturbation.

\end{document}